\documentclass[twocolumn,aps,prx,floatfix,showpacs,amsmath,amssymb]{revtex4-1}

\usepackage{epsfig}
\usepackage{graphicx}
\usepackage{dcolumn}
\usepackage{bm}
\usepackage[colorlinks=true,dvipdfm]{hyperref}

\begin{document}

\title{Nonequilibrium quantum criticality in open systems: The dissipation rate as an additional indispensable scaling variable}
\author{Shuai Yin} \email{zsuyinshuai@163.com}
\author{Peizhi Mai}
\author{Fan Zhong}  \email{stszf@mail.sysu.edu.cn}

\affiliation{State Key Laboratory of Optoelectronic Materials and
Technologies, School of Physics and Engineering, Sun Yat-sen
University, Guangzhou 510275, People's Republic of China}

\begin{abstract}
We propose that nonequilibrium quantum criticality in open systems at both zero and finite temperatures can be described by the theories of open quantum systems. A master equation of the Lindblad form is derived from a system coupling microscopic to a heat bath. It is suggested to be valid generally for studying dynamical quantum criticality and is thus designated as Model Q upon generalizing Hohenberg and Halperin's classification of classical critical dynamics. We find that the dissipation rate in the equation must be included in the scaling forms as an indispensable additional scaling variable in order to correctly describe the nonequilibrium quantum critical behavior. Yet, the equilibrium fixed point determines the nonequilibrium critical behavior in the weak coupling limit. Through numerically solving the Lindblad equation for the quantum Ising chain, we attest these propositions by finite-time scaling forms with the dissipation rate. Nonequilibrium dynamic critical behavior of spontaneous emissions in dissipative open systems at zero temperature near their quantum critical points is found and is also described well by the scaling forms. Model Q thus provides a general approach to study quantum critical behavior of a system itself through its weak coupling to the environment.
\end{abstract}

 \maketitle

\date{\today}

\section{Introduction}
Recent experiments in ultracold atoms including the realization of manipulating real-time evolutions \cite{Greiner,Kinoshita,Hofferberth,Zhang} have stimulated a lot of studies \cite{Dziarmaga,polrmp,Zurek} and improved our understanding of nonequilibrium behavior of quantum phase transitions (QPT) \cite{sachdev}. For example, the Kibble-Zurek mechanism (KZM) of an adiabatic--impulse--adiabatic approximation, first introduced in cosmology by Kibble \cite{kibble1} and then in condensed matter physics by Zurek \cite{zurek1} to describe the dynamics of classical phase transitions, has been generalized to characterize quantum criticality at zero temperature \cite{Dziarmaga,polrmp,Zurek,ZDziarmaga}. Full scaling forms have been proposed to describe the nonadiabatic behavior in the impulse region \cite{deng,dengde,dengko}. Finite time scaling (FTS) \cite{Gong,Zhong} provides an understanding of these scaling forms and convenient methods to determine critical properties such as the critical point and critical exponents \cite{Yin}. So far, nonequilibrium quantum critical behaviors (NQCBs), at least for driving dynamics of the KZM in closed systems, have been well described by the usual theory of critical phenomena without any new ingredient.

But is this true for NQCBs at finite temperatures at which all practical systems exist? The famous quantum critical region (QCR)~\cite{Chakravarty} exhibits a variety of exotic behavior in a wide range of strongly correlated systems~\cite{sachdev,Coleman,sachdevpt}. Thermal fluctuations and strongly correlated quantum entanglement make this issue difficult to tackle \cite{sachdev2}. Except for some special one-dimensional (1D) systems \cite{sachdev}, there is no analytic, semiclassical, or numerical methods of condensed matter physics to understand the dynamic quantum criticality at finite temperatures~\cite{sachdevpt}; let alone nonequilibrium effects. Current studies on the generalization of the KZM to finite temperatures can be divided into two catalogs. One is to start from a finite-temperature equilibrium initial state, drive the system out of equilibrium and then study its responses \cite{Polkovnikov,cardy,Gritsevp,dengor}. The other one, more experimentally practical, is to study a nonequilibrium system interacting with a heat bath~\cite{patane}. In this case, the density of excitations $n$ is approximated by $n_{\rm KZ}+n_{\rm inc}$. Here, $n_{\rm KZ}$ is produced in the absence of the bath and described by the KZM, while $n_{\rm inc}$ stems from the incoherent interaction with the bath in the QCR and is proportional to the coupling. It scales inversely with the quench rate and dominates for small rates. However, an approximate solution shows that $n$ saturates rather than diverges at vanishing rates~\cite{patane}. So, this `incoherent' scaling can only be a crossover. It is not clear whether the system--bath coupling is an independent scaling variable in the NQCB or just a proportional coefficient either. Further, can it affect the universal critical behavior?

Besides, driven dissipative open quantum systems at zero temperature have attracted much attention as they offer a promising route of quantum computations or state engineering~\cite{Verstraete}. Do the spontaneous emissions across a critical point of such a driven system exhibit scaling behavior and, if yes, is no new ingredient needed to describe it either?

To answer these questions, we shall study the universal critical behavior of a nonequilibrium quantum system interacting with a heat bath that defines its temperature. In the weak system--environment coupling limit, an explicit master equation of the Lindblad form~\cite{Lindblad,attal,Breuer,mai} is derived from the microscopic Hamiltonian. This Lindblad equation evolves to the equilibrium canonical distribution of the system itself. It also leads immediately to the conclusion that the dissipation rate, representing a system--environment interaction in the Lindblad equation, must be included as an independent scaling variable in order to describe the full scaling of the universal NQCB. Still, introduction of this new variable does not change the critical exponents as the equilibrium fixed point determines the universal NQCB in the weak coupling limit. These results are in sharp contrast with previous studies in which the dissipation is large enough to either change the equilibrium distribution to that of the combined system and environment~\cite{Wang} or induce a phase transition in the system even in the equilibrium situation~\cite{Sudip,Gerd,Weiss}. Owing to the quantum nature, the NQCB described by the Lindblad equation is also different from any model of classical critical dynamics classified by Hohenberg and Halperin~\cite{Hohenberg,Folk}. We therefore designate the Lindblad equation as Model Q for the NQCB in open systems.

The rest of paper is organized as follows: Section~\ref{slind} gives the derivation of the Lindblad equation and discuss some of its properties. Then in Sec. \ref{sgern}, we discuss generally the scaling behavior of the dissipation rate. To show explicitly the properties of the dissipation rate, we examine the scaling forms of FTS by numerically solving the Lindblad equation in Sec.~\ref{numfts}. Section~\ref{comp} compares our theory with previous scaling predictions and introduces the Q model with its main properties. A summary follows in Sec.~\ref{summ}.

\section{\label{slind}The Lindblad master equation}
In this section, we start with a generic microscopic Hamiltonian with a local interaction between a system and its environment and derive briefly a master equation of the Lindblad form under several usual approximations. This equation will then serve as the central equation for the subsequent study.

\subsection{Model and its diagonalization}
Consider the quantum transverse-field Ising model interacting with a bosonic heat bath~\cite{patane}. The total Hamiltonian is
\begin{equation}
\mathcal{H}=\mathcal{H}_S+\mathcal{H}_B+\mathcal{H}_{SB},\label{H}
\end{equation}
with
\begin{eqnarray}
\mathcal{H}_S=-h_x\sum_{i=1}^N\sigma_i^x-J\sum\limits_{i=1}^{N-1}\sigma_i^z\sigma_{i+1}^z,\label{Hamil1}
\end{eqnarray}
\begin{eqnarray}
\mathcal{H}_B=\sum_{\beta,i}\omega_\beta b_{\beta i}^\dagger b_{\beta i},\label{HB}
\end{eqnarray}
\begin{eqnarray}
\mathcal{H}_{SB}=\sum_{\beta,i}\lambda_\beta (b_{\beta i}^\dagger+b_{\beta i})\sigma_i^x,\label{HSB}
\end{eqnarray}
where $\mathcal{H}_S$ is the Hamiltonian for the Ising system that contains $N$ spins with the Pauli matrices $\sigma_i^x$ and $\sigma_i^z$ in a traverse field $h_x$, $\mathcal{H}_B$ and $\mathcal{H}_{SB}$ are the Hamiltonians for the heat bath and their interaction, respectively, $\lambda_\beta$ is the coupling strength, $b^\dagger_{\beta i}$ ($b_{\beta i}$) creates (annihilates) in mode $\beta$ with an energy $\omega_\beta$ a boson coupling to the spin at site $i$. In the absence of the heat bath, the Ising chain~(\ref{Hamil1}) exhibits a continuous QPT from a ferromagnetic phase to a quantum paramagnetic phase at the critical point $h_{xc}=1$ at $T=0$~\cite{sachdev} and is realized in CoNb$_2$O$_6$ experimentally~\cite{Coldea}. The order parameter of the transition is the magnetization $M=\sum_{n=1}^N\langle\sigma_n^z\rangle/N$ with the angle brackets denoting quantum and thermal averages.

In order to diagonalize $\mathcal{H}_S$, we first apply a Jordan-Wigner transformation to Eq.~(\ref{Hamil1}) and replace the spin operators with the fermion operators $C_{-q}^\dagger$ and $C_q$ in the momentum $q$ space. A Bogoliubov transformation to $\eta_q=u_q^*C_q+v_{-q}C_{-q}^\dagger$ with constants $u_q$ and $v_q$ then results in
\begin{eqnarray}
\mathcal{H}=\sum_q\Omega_q\eta_q^\dagger\eta_q+\sum_{\beta,q}\omega_q b_{\beta q}^\dagger b_{\beta q}+\sum_{j=1}^3\mathcal{H}_{jSB},
\end{eqnarray}
where $\Omega_q=2J\sqrt{(h_x/J)^2+1-2(h_x/J){\cos q}}$ is the energy spectrum of the transverse-field Ising model Eq.~(\ref{Hamil1}), $b_{\beta,q}$ is the Fourier transformation of $b_{\beta i}$, and $\mathcal{H}_{jSB}$ is given by
\begin{eqnarray}
\mathcal{H}_{jSB}=\frac {-1}{\sqrt{N}}\sum_{k,q,\beta}\lambda_\beta(b_{\beta j}^\dagger a_{jkq}+b_{\beta j}a_{jkq}^\dagger),
\end{eqnarray}
with
\begin{eqnarray}
a_{1kq}&=&(u_k^*u_{k+q}-v_{k+q}v_k^*)(\eta_k^\dagger\eta_{k+q}-\eta_{k+q}\eta_k^\dagger),\nonumber\\
a_{2kq}&=&(v_{-k}u_{k+q}-v_{k+q}u_{-k})\eta_{-k}\eta_{k+q},\nonumber\\
a_{3kq}&=&(u_k^*v_{-k-q}^*-u_{-k-q}^*v_k^*)\eta_k^\dagger\eta_{-k-q}^\dagger.\label{akq}
\end{eqnarray}
In the interaction picture whose variables are denoted by a superscript I, the interaction becomes
\begin{eqnarray}
\begin{split}
\mathcal{H}^I_{SB}(t)=&\frac {-1}{\sqrt{N}}\sum_{j=1}^{3}\sum_{k,q,\beta}\lambda_\beta\times \\& \left(b_{\beta j}^\dagger a_{jkq}e^{-i\omega_{jkq\beta}t}+b_{\beta j}a_{jkq}^\dagger e^{i\omega_{jkq\beta}t}\right),\label{18}
\end{split}
\end{eqnarray}
where $\omega_{jkq\beta}=\omega_{jkq}-\omega_\beta$ with $\omega_{1kq}=\Omega_{k+q}-\Omega_k$, $\omega_{2kq}=\Omega_{-k}+\Omega_{k+q}$, and $\omega_{3kq}=-\Omega_k-\Omega_{-k-q}$.

\subsection{Derivation of the Lindblad equation}
Evolution of the composite system is governed by the Liouville-von Neumann equation $d\rho/dt =-i[H,\rho]$, which reads in the interaction picture,
\begin{eqnarray}
\frac{\partial \rho^I(t)}{\partial t}=-i[\mathcal{H}^I_{SB}(t),\rho^I(t)].\label{Liouville}
\end{eqnarray}
where $\rho$ is the density matrix of the composite system. We have set the Planck constant $\hbar=1$ and shall let the Boltzmann constant $k_{\rm B}=1$ in the following. For a thermal equilibrium heat bath, its density matrix  $\rho_B^I=\textrm{exp}\left(-\mathcal{H}_B/T\right)/\textrm{Tr}_B\left[{\textrm{exp}\left(-\mathcal{H}_B/T\right)}\right]$, where $\textrm{Tr}_B$ is to trace the freedoms of the bath. In the weak system--bath interaction situation, $\rho^I(t)\simeq \rho_S^I(t)\otimes \rho_B^I$, where $\rho_S^I$ is the density matrix of the system. Under the Markovian approximation, which erasing the long time memories, the evolution of $\rho^I_S$ is then described by the Born--Markov equation~\cite{Breuer}
\begin{equation}
\frac{\partial\rho_S^I}{\partial t}=-\textrm{Tr}_B[\mathcal{H}_{SB}^I(t),\int\limits_0^{+\infty}dt'[\mathcal{H}_{SB}^I(t-t'),\rho_S^I(t)\otimes\rho_B^I]].
\end{equation}
Inserting Eq.~(\ref{18}) in this equation, using a rotating wave approximation~\cite{Breuer}, and neglecting the irrelevant lamb shift term~\cite{patane}, we find back in the Schr\"{o}dinger picture
\begin{eqnarray}
\begin{split}
\frac{\partial \rho_S}{\partial t}=&-i[\mathcal{H}_S,\rho_S]
-c\sum_{j=1}^2\sum_{k,q}[\gamma_{jkq}(\langle n(\omega_{jkq})\rangle+1)\times\\
&(\rho_S a_{jkq}^\dagger a_{jkq}+a_{jkq}^\dagger a_{jkq}\rho_S-2a_{jkq}\rho_S a_{jkq}^\dagger)]\\
&-c\sum_{j=1}^2\sum_{k,q}[\gamma_{jkq}(\langle n(\omega_{jkq})\rangle\times\\
&(\rho_Sa_{jkq}a_{jkq}^\dagger+a_{jkq}a_{jkq}^\dagger \rho_S-2a_{jkq}^\dagger \rho_Sa_{jkq})],\label{lindblad}
\end{split}
\end{eqnarray}
where $\gamma_{jkq}=\pi c_{\omega_{jkq}}/(cN)$, $c=\sum_\omega c_{\omega}/N'$ is the dissipation rate, $N'$ is the number of the total boson modes, and $c_{\omega}=\sum_\beta \lambda_\beta^2\delta_{\omega,\omega_\beta}$.

Although Eq.~(\ref{lindblad}) is obtained for a specific bosonic bath, it can be readily generalized to other baths. For two energy levels $E_l$ and $E_m$ with
$E_m-E_l=\omega_{jkq}$, we can make replacements of $V_{m\rightarrow l}\rightarrow a_{jkq}$ and
$V_{m\rightarrow l}^\dagger\rightarrow a^\dagger_{jkq}$, which are
quantum jump operators representing respectively emitting and absorbing a particle
with energy $\omega_{jkq}$ and jumping to a lower and higher energy
state. Also, let $W_{l\rightarrow m}\rightarrow\gamma_{jkq}\langle
n(\omega_{jkq})\rangle$ and $W_{m\rightarrow l}\rightarrow\gamma_{jkq}(\langle
n(\omega_{jkq})\rangle+1)$, the transition probabilities from
$l$th state to the $m$th and vice versa. We have
\begin{equation}
\frac{W_{l\rightarrow m}}{W_{m\rightarrow l}}=\frac{\gamma_{jkq}(\langle
n(\omega_{jkq})\rangle)}{\gamma_{jkq}(\langle
n(\omega_{jkq})\rangle+1)}=\frac{\textrm{exp}\left(-E_m/T\right)}{\textrm{exp}\left(-E_l/T\right)}.\label{c}
\end{equation}
Although the form of the transition probability
$W_{l\rightarrow m}$ depends on the environment and determines the details of the process, universal
properties only rely on $W_{l\rightarrow m}/W_{m\rightarrow l}$ constrained by Eq. (\ref{c}), which
is the detailed balance condition. After these replacements, Eq~(\ref{lindblad}) becomes
\begin{eqnarray}
\begin{split}
\frac{\partial \rho_S}{\partial
t}=&-i[\mathcal H_S,\rho_S]-c\sum_{m\neq l}W_{l\rightarrow m}(V_{l\rightarrow m}^\dagger V_{l\rightarrow m} \rho_S\\&+\rho_S V_{l\rightarrow m}^\dagger V_{l\rightarrow m}-2V_{l\rightarrow m} \rho_S V_{l\rightarrow m}^\dagger). \label{Lind}
\end{split}
\end{eqnarray}
This is the Lindblad equation that will be used to study the NQCBs in the following. It has been widely used in quantum optics \cite{orszag} and relaxation processes in open quantum systems \cite{attal,Breuer,znidaric}. However, it has not yet been applied to many-body systems to our knowledge.

\subsection{Special features}
We study here the equilibration properties of the Lindblad equation. The detailed balance condition Eq.~(\ref{c}) ensures that the steady solution of Eq.~(\ref{Lind}) for a time-independent $\mathcal H_S$ and its bath is the canonical distribution $\rho_E\equiv\textrm{exp}(-\mathcal H_S/T)/\textrm{Tr}
\textrm{exp}(-\mathcal H_S/T)$. Indeed, this condition and $V_{i\rightarrow j}=V_{j\rightarrow i}^\dagger$ nullify the second term on the righthand side of Eq.~(\ref{Lind}) for $\rho_S=\rho_E$, which satisfies $[\mathcal H_S,\rho_E]=0$. Note that this steady solution does \emph{not} depend on $c$ due to the weak coupling limit \cite{Schr,Goldstein,Cho}, a fact which gives a constraint to nonequilibrium scaling forms. 

Equation~(\ref{Lind}) includes both the quantum and thermal contributions. If the second term on the righthand side is neglected, it recovers the quantum Liouville
equation contributing quantum fluctuations to the evolution of $\rho_S$; while if the first term is omitted, the diagonal part of the remaining equation becomes
\begin{equation}
\frac{\partial \rho_{ii}}{\partial t}=c\sum_{j\neq
i}(W_{j\rightarrow i}\rho_{jj}-W_{i\rightarrow j}\rho_{ii})
\label{cmaster},
\end{equation}
where $\rho_{ii}$ is the probability of the system in the $i$th state. This is a master equation describing a classical stochastic process~\cite{Cardy}. Thus, Eq.~(\ref{Lind}) is a dynamical equation integrating both quantum and thermal contributions.

\section{\label{sgern}Scaling behavior of the dissipation rate}
In this section, we study the scaling properties of the dissipation in the Lindblad equation.

Equation~(\ref{Lind}) immediately indicates that one has to include $c$ as a new ingredient for the NQCBs, as it resembles formally dynamic equations of classical critical dynamics with mode couplings~\cite{Hohenberg,Folk} though for $\rho_S$ instead of an order parameter. Physically, it is clear that the system--bath interaction, whose effect reduces to $c$, can influence quantum critical behavior of the system. The stronger the interaction with the environment is, the faster the system equilibrates, and so the shorter the relaxation time becomes. The reversal Liouville part of the equation specifies the quantum nature of the dynamics and underlies the distinctive features of the open quantum systems. It cannot thus be omitted; otherwise the dynamics would return completely to classical one governed by Eq.~(\ref{cmaster}). As this part has defined the microscopic time scale for quantum fluctuations, $c$ may be regarded as a kind of a coupling instead of simply a kinetic coefficient for a time scale and cannot be eliminated by rescaling.

We now argue that $c=0$ is the relevant fixed point for NQCBs in the weak coupling limit. Note first that the dimension of $c$ is identical with $t^{-1}$ as can be seen from Eq.~(\ref{Lind}) by dividing it by $c$. This means that $c$ is relevant. However, it is only a ``dynamically relevant'' variable that influences only nonequilibrium behavior. This is because the dissipation only manifests in nonequilibrium situations where energy exchanges with the bath are needed for the system to equilibrate; in equilibrium, it has no effect at all as the equilibrium solution of Eq.~(\ref{Lind}) is always $\rho_E$, independent of $c$ in the weak coupling limit as mentioned above. So, although $c$ would grow up upon coarse graining in dynamics if it did not initially lie at a fixed point, in equilibrium, the scaling behavior including the critical exponents must recover the equilibrium one. For large couplings, the equilibrium distribution may contain both the system and the bath together and may be governed by a fixed point with a finite $c$~\cite{Wang,Sudip,Gerd,Weiss}. However, this is beyond the applicability of the Lindblad equation. So, in the weak coupling regime, a reasonable fixed point should then be $c=0$, which is just the fixed point that controls the equilibrium behavior including the scaling of $c$ near it. This is similar to the Ising universality class in the classical critical phenomena in which the magnetic field is a relevant perturbation. A difference is that the magnetic field has a new exponent but here no new one for $c$ is necessary due to its dimension. We shall show in the following section that these are true.

Generally, there may exist three kinds of possible fixed points for the dissipation in a system with a continuous phase transition under the renormalization-group transformation. The first kind corresponds to an infinite dissipation. In this case, quantum fluctuations are negligible and the dissipation controls completely the phase transition. The dynamics is described by Eq.~(\ref{cmaster}) and thus is classical. The second one corresponds to a finite dissipation. Both the quantum and the dissipative terms contribute and a new phase transition different from that of the system alone may be induced. This is the so-called dissipative phase transition~\cite{Sudip,Gerd,Weiss}. The third kind of fixed points corresponds to a zero dissipation. This is just the case here in which the phase transition is controlled by the original system Hamiltonian and so is still the quantum phase transition. However, because the dissipation is relevant, we must thus include it as an indispensable scaling variable in order to describe correctly the quantum critical behavior.

We note that in the case of the dissipative phase transition~\cite{Sudip,Gerd,Weiss}, the dissipation is so large that it can push effectively the system to a new phase. Thus a new fixed point with a finite dissipation emerges to control the dissipation-induced phase transition, if this phase transition is continuous. This occurs even in the equilibrium situation. In the present case on the other hand, the dissipation is too weak to change the equilibrium properties, even if it may well flow to the finite fixed point. In other words, the dissipation is only a relevant perturbation near the original fixed point. Moreover, as mentioned above, the dissipation rate manifests itself only in the nonequilibrium situation. This is also different from the magnetic field which shows up even in equilibrium. We suggest that the Lindblad equation~(\ref{Lind}) is just suitable for describing the dynamic quantum critical behavior in open systems in which the dissipation is a small relevant perturbation.

\section{\label{numfts}Verification of indispensability of the dissipation rate via FTS}

\subsection{FTS}
In order to verify explicitly the theory proposed, we employ the theory of FTS by applying a linearly sweeping field to drive the system across its critical point at a rate $R$.
Such a driving imposes an external time scale on the system to manipulate its evolution from $t\rightarrow -\infty$. This time scale is $\tau_d\sim R^{-z/r}$~\cite{Gong,Zhong,Yin}. FTS~\cite{Gong,Zhong,Yin} provides a unified theory to understand the nonequilibrium critical behavior arising from the competition of $\tau_d$ with other time scales including the intrinsic one $\tau_s\sim |g|^{-\nu z}$, where $g=h_x-h_{xc}$ and $\nu$ is the correlation length critical exponent. In particular, when $\tau_d<\tau_s$, which happens in the vicinity of the critical point $g=0$, the system falls out of equilibrium. The scaling behavior of such a driven system can be readily obtained from the scale transformation,
\begin{widetext}
\begin{equation}
M(g,h_z,T,L,R,c,t)= b^{-\beta /\nu}M(gb^{1/\nu}, h_zb^{\beta\delta/\nu}, Tb^{z}, Lb^{-1},Rb^r,cb^{z}, tb^{-z}),\label{FTSn2}
\end{equation}
for the order parameter $M$, where $L$ is the size of the system, and $\beta$ and $\delta$ are critical exponents defined by $M\propto g^\beta$ when the symmetry breaking field $h_z$ conjugate to $M$ vanishes and $M\propto h_z^{1/\delta}$ at the critical point~\cite{Cardy}. We have neglected dimensional factors for simplicity. We emphasize that in the classical case, simple inclusion of the coarse-graining time $tb^{-z}$ generalizes critical dynamics~\cite{Hohenberg} directly to nonequilibrium~\cite{Gong}. However, in the quantum nonequilibrium process described by Eq.~(\ref{Lind}), $c$ must be included as an additional variable.

The scale transformation Eq.~(\ref{FTSn2}) is an extension of the usual one both in classical critical phenomena~\cite{Cardy} and in quantum phase transitions~\cite{sachdev} to include the additional $R$. It provides a simple method to derive critical behavior. For example, if we just keep the first two arguments of the order parameter as in the case of equilibrium critical phenomena, choosing $b\propto g^{-\nu}$ so that the first argument on the left is a constant then results in $M=g^{\beta}f(h_zg^{-\beta\delta})$ ($f$ is a scaling function), which just defines the critical exponent $\beta$ at $h_z=0$. Similarly, choosing $b\propto h_z^{-\nu/\beta\delta}$ at $g=0$ leads to $M\propto h_z^{1/\delta}$. In addition, assume that a variable $\lambda$ has a dimension $[\lambda]$ and $\lambda=0$ corresponds to a critical point. If $\lambda=Rt$, the scale transformation gives $\lambda b^{-[\lambda]}=R b^r tb^{-z}$. As a result, $r=-[\lambda]+z$.

Various kinds of driving dynamics can be defined from Eq.~(\ref{FTSn2}).
If we choose $\lambda$ as the external field in the longitudinal direction, i.e., $h_z=R_zt$, Eq.~(\ref{FTSn2}) becomes
\begin{equation}
M_h(g,h_z,T,L,R_z,c)=R_z^{\beta /\nu r_z}f_1(gR_z^{-1/\nu
r_z},
h_zR_z^{-\beta \delta /\nu
r_z},TR_z^{-z/r_z},L^{-1}R_z^{-1/r_z},cR_z^{-z/r_z})\label{op2}
\end{equation}
where $r_z=z+\beta\delta/\nu$ and $f_i$ for integer $i$ is a scaling function. We have suppressed a redundant variable among the trio $h_z$, $R_z$, and $t$. Of course, we can vary $g$ as $g=R_xt$. Then, Eq.~(\ref{FTSn2}) leads to the FTS form
\begin{equation}
M_g(g,h_z,T,L,R_x,c)= R_x^{\beta /\nu r_x}
f_2(gR_x^{-1/\nu
r_x}, h_zR_x^{-\beta\delta/\nu r_x},TR_x^{-z/r_x},L^{-1}R_x^{-1/r_x},cR_x^{-z/r_x}) \label{opg}
\end{equation}
with $r_x=z+1/\nu$. For a closed system $c=0$ in the thermodynamic limit $L\rightarrow\infty$ and at $T=0$, this driving recovers the ordinary KZM and has been postulated in Ref.~\cite{deng}. Further, we can cool linearly the system to the quantum critical point via $T=-R_Tt$ and obtain
\begin{equation}
M_T(g,h_z,T,L,R_T,c)=R_T^{\beta /\nu r_T} f_3(gR_T^{-1/\nu
r_T},h_zR_T^{-\beta \delta /\nu
r_T},TR_T^{-z/r_T},L^{-1}R_T^{-1/r_T},cR^{-z/r_T}), \label{opT}
\end{equation}
\end{widetext}
with $r_T=2z$ from Eq.~(\ref{FTSn2}). Sweeping temperature can be readily realized in cold atom experiments \cite{David,phillips}.

\subsection{Model and numerical method}
In order to verify the scaling behavior, we directly solve Eq.~(\ref{Lind}) numerically for the Ising chain (\ref{Hamil1}) with an additional small symmetry-breaking term $-h_z\sum_{i=1}^N \sigma_i^z$ and $W_{j\rightarrow i}=\beta_i$. For this quantum Ising chain, $\beta=1/8$, $\delta=15$, and $\nu=z=1$.
A finite difference method to second order is used. The lattice sizes used are $L=6$, $7$, and $8$ with periodic boundary conditions. Although these sizes appear small, we shall see that the finite-size scaling as has been taken into  account in the last section still works well. The time interval is $0.0005$. Smaller values were checked to produce no appreciable changes.

We note that the Lindblad operator in Eq.~(\ref{Lind}) has been written in a matrix form, which, as inherits from quantum mechanics, contains all the eigenstates in principle~\cite{Breuer} and may thus be difficult to be solved. An existing numerical method~\cite{Verstraete1} considers only a local dissipation operator, while an analytic method~\cite{Prosen} needs a quadratic form of the Hamiltonian. Nevertheless, it may be solved numerically by employing recent numerical renormalization-group methods~\cite{Verstraete1} to eliminate the least relevant ones. Usually only a few excited states contribute for slow driving, since states with $E_i\gg T$ play negligible roles as their contributions to $\rho_E$ decrease exponentially. Here for the sake of demonstration, we are content with the direct method.

\subsection{Numerical results}
\begin{figure}
\centerline{\epsfig{file= 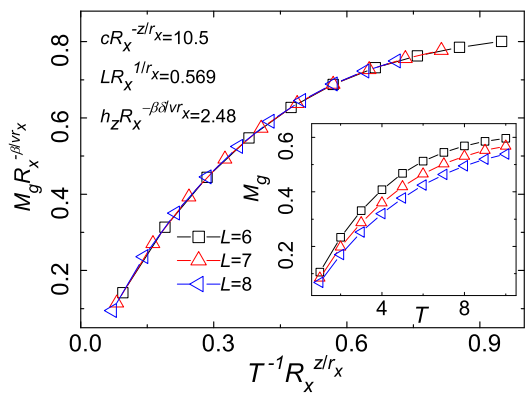,width=1.0\columnwidth}}
\caption{\label{cpg} (color online) $M_g$ versus $T$ plotted for different lattice sizes in the inset collapse onto each other after rescaling.}
\end{figure}

We first examine the scaling form~(\ref{opg}) for varying $g$, a usual Kibble-Zurek protocol~\cite{patane}. As it contains several variables, we fix some of them. In Fig.~{\ref{cpg}} we plot the curves of $M_g$ versus $T$ at $g=0$ for fixed $h_zR^{-\beta\delta/\nu r_x}$, $LR^{1/r_x}$ and $cR^{-z/r_x}$. The rescaled curves collapse well, which confirms Eq.~(\ref{opg}). This indicates that the KZM must be modified to include the dissipation effects when the quantum systems are open.

\begin{figure}
\vskip -1.8cm
  \centerline{\epsfig{file= 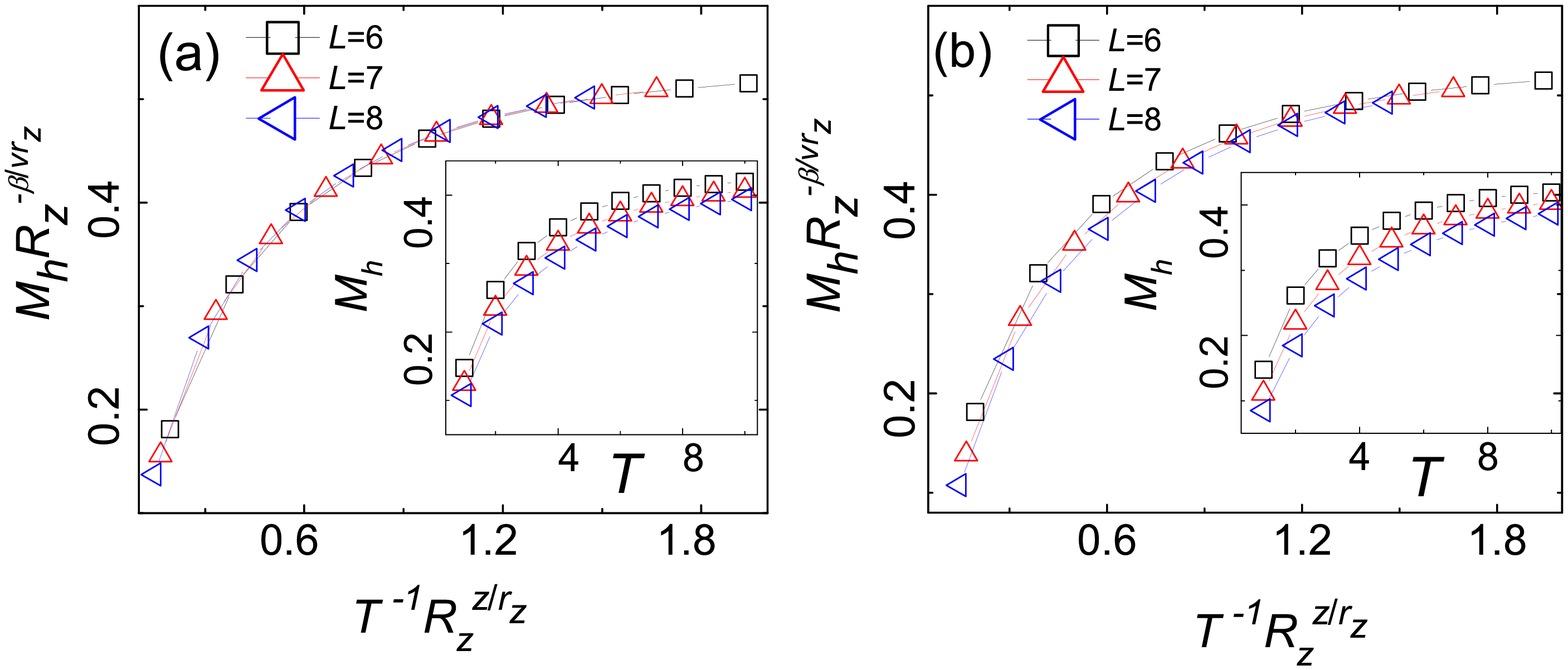,width=1.0\columnwidth}}
  \caption{\label{cp1} (a) At fixed $LR_z^{1/r_z}=1.17$ and $cR_z^{-z/r_z}=0.515$, the curves of $M_h$ versus $T$ plotted in the inset for several lattice sizes collapse onto each other after rescaling as expected from Eq.~(\ref{op2}). (b) At fixed $LR_z^{1/r_z}=1.17$
   and $c=0.100$ instead, the curves separate from each other. Thus the dissipation rate is indispensable in the dynamic scaling forms at finite temperatures.}
\end{figure}

\begin{figure}
  \centerline{\epsfig{file= 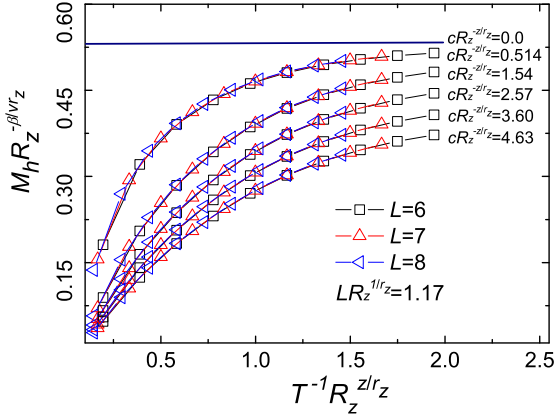,width=1.0\columnwidth}}
  \caption{\label{cp2} (color online) $M_hR_z^{-\beta/\nu r_z}$ versus $T^{-1}R_z^{z/r_z}$ at fixed $LR_z^{1/r_z}$ and various $cR_z^{-z/r_z}$. These rescaled curves approach a constant (the horizontal line) at $c=0$ as $cR_z^{-z/r_z}$ decreases.}
\end{figure}

\begin{figure}\centerline{\epsfig{file= 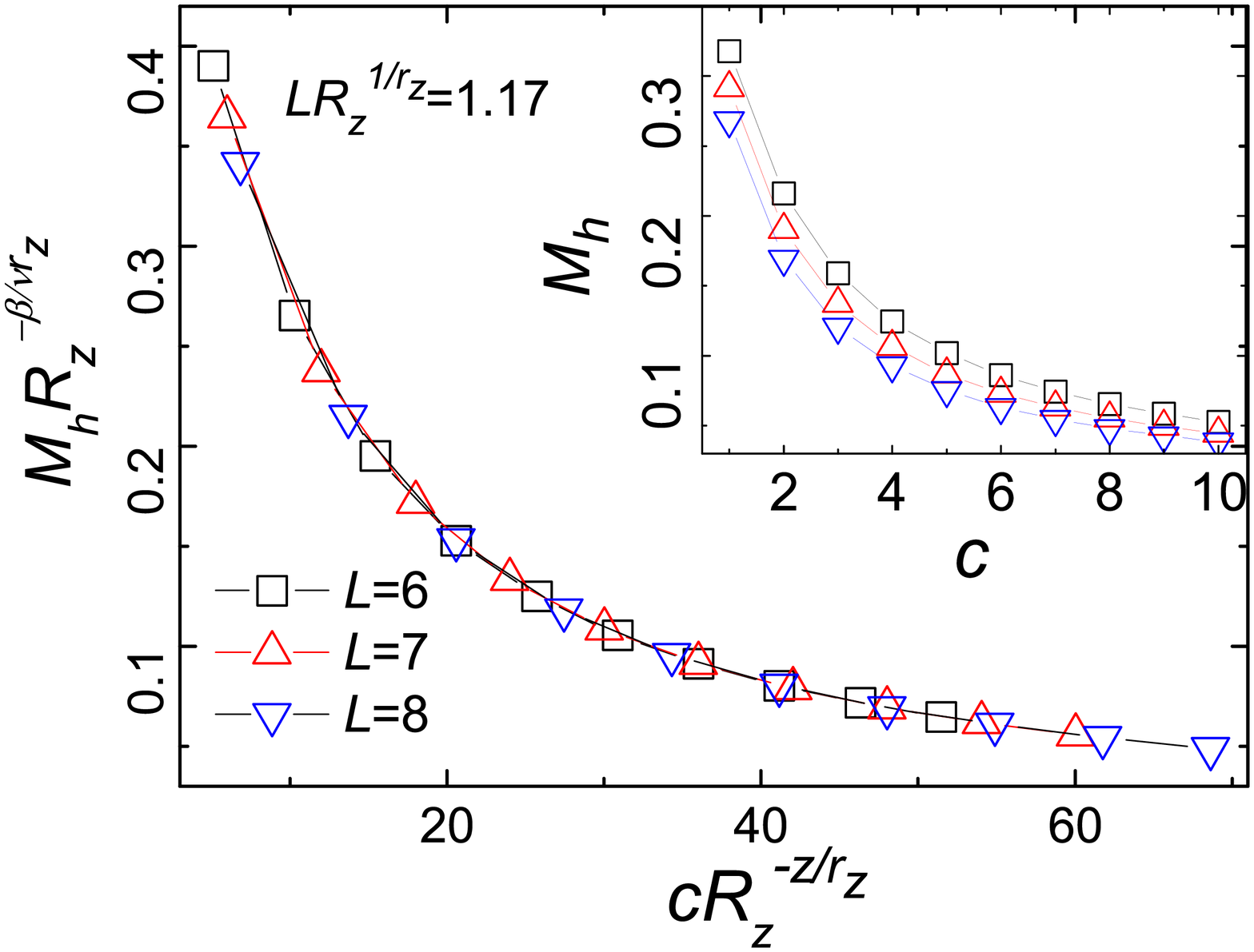,width=1.0\columnwidth}} \caption{\label{cp3} Critical behavior of spontaneous emissions at $T\rightarrow0$. $M_h$ versus $c$ plotted for several lattice sizes and fixed $LR_z^{1/r_z}$ in the inset collapse after properly rescaled.}\end{figure}

Next, we focus on Eq.~(\ref{op2}) for varying $h_z$. Again, upon fixing some variables, the resultant excellent scaling collapses depicted in Figs.~\ref{cp1}(a) confirm the validity of the scaling form. Figure~{\ref{cp1}}(b) shows clearly that if $c$ is not properly rescaled according to Eq.~(\ref{op2}), the rescaled curves separate from each other even though $c$ is small. This may provide a hint at the additional variable for experiments. Figure~\ref{cp2} plots the rescaled curves for various fixed values of $cR_z^{-z/r_z}$. Although $c$ varies from $0.1$ to $0.9$ for $L=6$, Eq.~(\ref{op2}) characterizes the scaling perfectly. For $T\rightarrow 0$, Eq.~(\ref{Lind}) describes a dissipative open quantum system that relaxes to its ground state by spontaneous emissions. The latter must thus obey~(\ref{op2}) too. This is confirmed in Fig.~{\ref{cp3}} for fixed $LR_z^{1/r_z}$. Therefore, $c$ is necessary and the equilibrium fixed point indeed controls the nonequilibrium behavior.

Finally, we verify the scaling form~(\ref{opT}). At fixed $cR^{-z/r_T}$, $h_zR_T^{-\beta \delta /\nu r_T}$, $LR_T^{1/r_T}$, and $g=0$, the curves of $M_T$ versus $T$ collapse onto each other according to Eq.~(\ref{opT}) as clearly demonstrated in Fig.~\ref{cp3}. Note that varying $T$ alters directly $W_{j\rightarrow i}$ and thus the last term of Eq.~(\ref{Lind}), the contribution of thermal fluctuations, in contrast to varying $h_z$ which makes $\mathcal H_S$ time dependent. Here, the system may be envisioned as interacting with a series of baths at different temperatures~\cite{patane2}. Then the dissipation rate $c$ does bring the information of changing the temperature of the bath to the system. Thus $c$ plays a crucial role in this process.
\begin{figure}
\centerline{\epsfig{file= 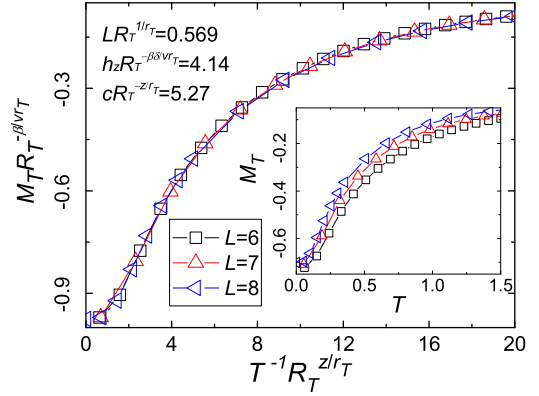,width=1.0\columnwidth}}
\caption{\label{cp5} (color online) $M_T$ versus $T$ plotted for several lattice sizes collapse after rescaled.}
\end{figure}

\section{\label{comp}Comparisons, Model Q and its possible extension}
In this section, we compare our general theory with the specific results in~\cite{patane}. In~\cite{patane}, only sweeping $g$ was considered. The defect density in the process up to the time $t_{\rm QC}\sim T^{1/\nu z}/R_x$ for $cT^{1/\nu z}\ll R_x$ was supposed to be $n\sim n_{\rm KZ}+n_{\rm inc}\sim R_x^{d/r_x}+cT^{(d\nu+1)/\nu z}/R_x$, which is
\begin{equation}
n\sim R_x^{d/r_x}[1+(cR_x^{-z/r_x})(TR_x^{-z/r_x})^{d/z+1/\nu z}]\label{np}
\end{equation}
if we assume $c$ corresponds to the relaxation rate $\tau^{-1}$ used in~\cite{patane} by comparing the dynamic equations, though the former is $T$ independent. In the present theory, similar to Eq.~(\ref{opg}) at $L\rightarrow\infty$ and $h_z=0$, we have
\begin{equation}
n= R_x^{d/r_x}f_4(gR_x^{-1/\nu r_x}, cR_x^{-z/r_x}, TR_x^{-z/r_x})\label{nfts}
\end{equation}
 for $T\ll R_x^{z/r_x}$ in the FTS domain~\cite{Yin}. For $T\gg R_x^{z/r_x}$ in the QCR, it crossovers to
\begin{equation}
n=T^{d/z}f_5(gT^{-1/\nu z}, cT^{-1}, R_xT^{-r_x/z})\label{nqcr}.
\end{equation}
At $t_{\rm QC}$ at which $g_{\rm QC}\equiv R_xt_{\rm QC}\sim T^{1/\nu r_x}$, $g_{\rm QC}T^{-1/\nu r_x}\sim1$ and $g_{\rm QC}R_x^{-1/\nu r_x}\sim (TR_x^{-z/r_x})^{1/\nu z}$ where another crossover happens. So, $f_4(u,v,w)=f_6(v,w)$. We see therefore that the two variables within the parentheses in Eq.~(\ref{np}) are just the scaling variables of $f_6$ and thus Eqs.~(\ref{np}) and (\ref{nfts}) agree qualitatively. Moreover, if we assume that $f_6(v,w)\sim {\rm constant}+vw^{(d\nu+1)/\nu z}$ near the crossover, we obtain Eq.~(\ref{np}) explicitly. Further, the approximate solution for the quantum Ising chain up to $t_{\rm QC}$ is $n_{\rm inc}\sim T(1-e^{-2cT/R_x})$~\cite{patane}. When $R_x\rightarrow 0$, $n_{\rm inc}\sim T^{d/z}$ in agreement with Eq.~(\ref{nqcr}). In addition, the exponent of the solution can be rewritten as a ratio of the two variables $cT^{-1}$ and $R_xT^{-r_x/z}$, again in consistence with our theory.

On the other hand, when $T\rightarrow 0$, $n_{\rm inc}\rightarrow0$ and there would be no incoherent contribution. However, our results show that the dissipation gives rise to spontaneous emissions that scale according to Eq.~(\ref{nfts}).

In these comparisons, we have replaced the relaxation rate $\tau^{-1}$ and its temperature dependence with the dissipation rate $c$. In this way, no details of the bath except its temperature appear in the scaling behaviors and universality is unveiled. For comparison, $\tau^{-1}\propto T^{s+d/z}$ with $s$ characterizing the bath spectral density ($s=1$ for an Ohmic bath)~\cite{patane}. Accordingly, the scaling would depend on the details of the bath.

Classical critical dynamics has been well classified by Hohenberg and Halperin~\cite{Hohenberg}. In Sec.~\ref{sgern}, we have taken advantage of this scheme and arrived at the relevancy of $c$. Questions then arise naturally as to which universality class does the present case belong? Or, can the quantum critical dynamics be classified according to this scheme? Does it need a new classification scheme? As we pointed out above, the Lindblad equation~(\ref{Lind}) includes both the quantum and thermal fluctuations, which applies at least to low temperatures and weak dissipations. For the quantum Ising chain studied, the dynamic critical exponent $z=1$, when the quantum nature cannot be ignored. This fact alone makes the model distinct to all previous classified catalogs. If we still stick to the classical classification, however, the Lindblad model~(\ref{Lind}) belongs to a new class. Accordingly, we break the naming sequence and call it Model Q for its quantum nature.

For this Model Q, a closed system at $T=0$ corresponds to $c=0$ and Eq.~(\ref{Lind}) describes just a continuous quantum phase transition of such a system. The quantum phase transitions of open systems with weak coupling to heat baths are controlled by the $c=0$ fixed point. Its infinite dissipation fixed point mentioned in Sec.~\ref{sgern} leads to the classical Model A and possibly others depending on the nondissipative couplings with other slow modes in the Hamiltonian~\cite{Hohenberg}. The finite dissipation fixed point that gives rise to dissipative phase transitions is not, however, described by the Q model, as Eq.~(\ref{Lind}) only possesses an equilibrium distribution of the original Hamiltonian itself. Nevertheless, the model offers an approach to study the quantum critical dynamics of a system itself at both finite temperatures and zero temperature.

We have followed the classical classification and defined Model Q. However, there is an important difference between classical and quantum critical behavior. In classical critical phenomena, statics and dynamics are decoupled. But in quantum phase transitions, they cannot be separated~\cite{sachdev}. Accordingly, the classification on the basis of the sole dynamic properties does not work for continuous quantum phase transitions in closed systems at $T=0$. Still, it is quite appealing whether or not there exist a quantum counterpart of the classical models with a primary or secondary conserved field~\cite{Hohenberg,Folk}. This may be called Model X if exists.

\section{\label{summ}Summary}
By definition, a quantum phase transition happens at zero temperature. However, it is now well known that the zero-temperature quantum critical point governs the behavior in the QCR at finite temperatures. Yet it is extremely hard to tackle the dynamics in this region, a problem which has a pivotal role in understanding the complex behavior. In experiments, a system obtains its temperature through a heat bath, no matter whether it is equilibrium or not~\cite{note}. So, an open system is an appropriate starting point. Here, we have proposed a Lindblad master equation that includes both the thermal and quantum fluctuations as Model Q to study the dynamic quantum criticality of such an open system. We have shown that this provides a valuable approach to study quantum critical dynamics of a system itself, as its equilibrium fixed point controls the nonequilibrium critical behavior in the weak system--bath coupling limit. Moreover, we have found that the dissipation rate must be included as a new scaling variables in order to correctly describe the universal dynamic quantum scaling behavior. This is also true for the dynamic critical behavior revealed in the spontaneous emissions of a dissipative open system at zero temperature near its quantum critical point. For demonstration, we have utilized the quantum Ising chain and FTS as examples. Yet, the theory should apply to other quantum critical systems as well.

We have shown clearly that the dissipation rate must be included in the nonequilibrium quantum critical behavior. It stems from the coupling to the heat bath. Experimentally, the heat bath does give a system its temperatures. But it is intrigued how they conspires to create a parameter of the dissipation rate other than the temperature. We are seeking answers.

\section*{Acknowledgement}
We thank Chaohong Lee and Junhong An for their helpful discussions. We also thank Xizhou Qin for his help in improving some Fortran codes. This project was supported by NNSFC (10625420) and FRFCUC.


\begin{thebibliography}{99}
\bibitem{note}Another method is to define the temperature of a nonequilibrium system as its initial equilibrium value. As a consequence, the scaling relation depends on the initial distribution. See [6] for an example.
\end{thebibliography}
\end{document}